\documentclass[preprint,12pt]{elsarticle}

\usepackage{amssymb}
\usepackage{amsmath}
\usepackage{siunitx}
\usepackage{booktabs}
\usepackage{mathrsfs,slashed}
\usepackage{nicefrac}
\usepackage{colorprofiles}

\usepackage{hyperref}

\usepackage[ngerman,english]{babel}

\usepackage{xcolor}

\journal{Physics Letters B}

\begin{document}

\begin{frontmatter}

\title{Cluster production and the chemical freeze-out in expanding hot dense matter}

\author[label1,label2,label3]{D.~Blaschke} 

\affiliation[label1]{Institute of Theoretical Physics, University of Wroclaw, Max Born pl. 9, 50-204 Wroclaw, Poland}
\affiliation[label2]{Center for Advanced Systems Understanding (CASUS), D-02826 Görlitz, Germany}
\affiliation[label3]{Helmholtz-Zentrum Dresden-Rossendorf, D-01328 Dresden, Germany}

\author[label4]{S.~Liebing}

\affiliation[label4]{TU Bergakademie Freiberg, Institute of Theoretical Physics, Leipziger Str. 23, 09599 Freiberg, Germany}
\author[label5,label1]{G.~R\"opke}
\affiliation[label5]{University of Rostock, Albert-Einstein-Str. 23-24, 18051 Rostock, Germany}
\author[label6]{B.~D\"onigus}
\affiliation[label6]{Institut für Kernphysik, Goethe-Universität Frankfurt, Max-von-Laue-Str. 1, 60438 Frankfurt, Germany}

\begin{abstract}
We discuss medium effects on light cluster production in the QCD phase diagram 
by relating Mott transition lines to those for chemical freeze-out.
In heavy-ion collisions at highest energies provided by the LHC, light cluster abundances should follow the statistical model because of low baryon densities. Chemical freeze-out in this domain is correlated with the QCD crossover transition. 
At low energies, in the nuclear fragmentation region, where the freeze-out interferes with the liquid-gas phase transition, self-energy and Pauli blocking effects are important.  
We demonstrate that at intermediate energies the chemical freeze-out line correlates with the maximum mass fraction of nuclear bound states, in particular $\alpha$ particles.
In this domain, the HADES, FAIR and NICA experiments can give new insights. 
\end{abstract}




\end{frontmatter}

\section{Introduction}

A characteristic result of heavy-ion collisions (HIC) are the yields of the observed clusters. 
Already from the first measurements of the yields of isotopes ($n$, $p$, $^2$H, $^3$H, $^3$He, $^4$He ($\alpha$), etc.) in relativistic heavy-ion collisions at Bevalac energies \cite{Gutbrod:1976zzr,Gosset:1976cy,Westfall:1976fu} 
it was a surprising result that these distributions of the yields are similar to the mass action law known from chemical equilibrium of a gas of reacting components, which is given by a temperature $T$ and a chemical potential $\mu$ for the baryon number
of the elementary constituents.
The so-called statistical model assumes that in HIC hot and dense matter (fireball) is produced
where the interaction within this dense system is strong so that the relaxation time to establish equilibrium is very short~\cite{Andronic:2017pug}.
We can assume local thermodynamic equilibrium at this stage.

This hot and dense matter expands, and the reaction rates become small so that chemical equilibrium is no longer established.
At the time $t_{\rm cf}$ the chemical composition freezes out. 
The temperature $T_{\rm cf}$ and chemical potential $\mu_{\rm cf}$ are parameter values which determine in this picture the primordial distributions that evolve after freeze-out to the observed yields. 
It was also surprising that with the advent of ultrarelativistic HIC at CERN and Brookhaven National Laboratory the description of the yields of hadrons as well as light nuclei within the statistical model with just two parameters ($T_{\rm cf},\mu_{\rm cf}$) for 
the chemical freeze-out worked remarkably well
\cite{Andronic:2017pug,Vitiuk:2020jsa,Bugaev:2020sgz}.
For the evolution of these sets of freeze-out parameters as a function of the center of mass energy of the collision, fit formulas for the behaviour of
$T_{\rm cf}(\mu)$ have been parametrized, e.g., by Cleymans {\it et al.} \cite{Cleymans:2005xv}, Vovchenko {\it et al.} \cite{Vovchenko:2015idt}, Andronic {\it et al.} \cite{Andronic:2017pug} and Poberezhniuk {\it et al.} \cite{Poberezhnyuk:2019pxs}. 

This simple approach has to be replaced by a more fundamental approach to nonequilibrium processes.
A discussion has been performed whether chemical 
freeze-out could be understood analogously to kinetic freeze-out as a reaction-kinetic process, when chemical reaction rates equal the expansion rate of the fireball, see
\cite{Heinz:2007in} and references therein.
A sufficiently strong medium dependence of the chemical reaction rates at the hadronization transition has been obtained in \cite{Blaschke:2011hm,Blaschke:2011ry} by assuming a geometrical hadron-hadron cross section \cite{Povh:1987ju} with medium dependent hadron radii that get strongly enlarged at the Mott dissociation transition, as demonstrated in the Nambu-Jona-Lasinio model for case of the pion at finite temperature \cite{Hippe:1995hu}. 
Along these lines of argument, the chemical freeze-out of hadrons from the quark-gluon plasma can be understood due to the dramatic localization of the hadronic wave functions triggered by the chiral symmetry breaking transition that leads to the binding of quarks in hadrons (reverse Mott dissociation of hadrons).

The task of understanding the evolution of particle distributions towards chemical freeze-out can be solved using the method of nonequilibrium statistical operator (NSO) as outlined in the following section. 
In addition, the use of the local thermodynamic equilibrium to describe the system at freeze-out allows to account for correlations in the system, i.e. bound states and continuum correlations.
The treatment of these correlations is the main issue of the present work.

An alternative approach to describe the nonequilibrium evolution of the expanding fireball are transport equations \cite{TMEP:2022xjg}, where cluster production is addressed by quantum molecular dynamics \cite{Glassel:2021rod} or kinetic approaches \cite{Kireyeu:2024woo}.
These kinetic equations describe the time evolution of the single-particle distribution function.
However, it is difficult to include the formation of correlations and bound states in a systematic way. 
In principle, the method of the nonequilibrium statistical operator is able to give a systematic approach, but this has been performed until now only in first steps \cite{Ropke:1988ymk,Danielewicz:1991dh,Kuhrts:2000zs,Kanada-Enyo:2012yif}.

In this work we start from a general approach to nonequilibrium which includes both the hydrodynamic approach and the kinetic approach. 
The freeze-out concept is connected with a quick change in the relevant parameters, for instance the change of density during a phase transition.
New observables become relevant such as the distribution function of elementary particles and clusters.
The chemical evolution has to be described by a reaction network, and one has to consider the feed-down by the decay of excited states which transform the primordial equilibrium distribution at freeze-out to the final distribution of yields as observed in the experiments.

As a main result, in addition to the hadronization phase transition at high temperatures (\SI{156}{MeV}~\cite{Andronic:2017pug,Gross:2022hyw}) and the nuclear matter phase transition below the critical temperature\footnote{There are different results for the critical point of the n.m. phase transitions. Mean-field approximations for the equation of state give higher values for the critical temperature in the range of \SIrange{14}{18}{\MeV}\,\cite{Typel:2009sy}.}  of about 
$T_{\rm liquid-gas}\sim$\SI{12.1}{MeV}~\cite{Typel:2009sy}, 
we consider the  formation of nuclear bound states (clusters) with mass number $A$ and proton number $Z$ at the Mott density $n_{A,Z}^{\rm Mott}(T)$ as a relevant process which determines the freeze-out in the range between the high- and low-temperature phase transitions.
A comparison with empirical values for $T_{\rm cf}$ is made.
While the chemical potential $\mu_{\rm cf}(T)$ at freeze out is well described, the baryon number density $n(T,\mu_{\rm cf}(T))$ is more involved and its evaluation requires the account of bound state formation including in-medium effects.

\section{The method of nonequilibrium statistical operator}

Using the method of the nonequilibrium statistical operator 
(NSO) \cite{Zubarev}, a nonequilibrium process is described by the statistical operator $\rho(t)$,
\begin{equation}
\label{rhoZ}
\rho(t)=\lim_{\epsilon \to 0} \epsilon \int_{-\infty}^{ t} d t' e^{-\epsilon ( t- t')}
e^{-\frac{i}{\hbar} H ( t- t')}\rho_{\rm rel}( t')e^{\frac{i}{\hbar} H ( t- t')}~,
\end{equation}
which is a solution of the von Neumann equation 
with boundary conditions $\langle B_i \rangle^{t'}$, $t' \le t$,
characterizing the state of the system in the past. This information is contained in the relevant statistical operator $\rho_{\rm rel}(t')$
which is constructed from the maximum of information entropy at given averages of relevant observables $B_i$. 
We introduce Lagrange multipliers $\lambda_i(t')$ and obtain a generalized Gibbs distribution for $\rho_{\rm rel}(t')$. 
The elimination of the Lagrange multipliers by the boundary conditions leads to the nonequilibrium generalization of the equations of state.

For instance, in nuclear matter a set of relevant observables are the  energy $H$ as well as the particle numbers $N_n, N_p$
of neutrons and protons, respectively, which are conserved quantities when weak processes are not considered.
The solution for the maximum entropy is the generalized Gibbs distribution 
 $\rho_{\rm rel}(t') \propto \exp[-(H-\lambda_n(t') N_n-\lambda_p(t') N_p)/\lambda_T(t')]$. 
These Lagrange multipliers $\lambda_i(t')$ (which are in general time dependent and for the hydrodynamical approach also depending on position), 
are not identical with the equilibrium parameters $T$ and  $\mu_\tau$, $\tau =n,p$, but may be considered as nonequilibrium generalizations 
of  temperature and chemical potentials. 

The NSO allows the possibility of including other observables if they become relevant for the characterization of the nonequilibrium state.
To derive kinetic equations, the occupation numbers of the quasiparticle states should be included into the set of relevant observables $B_i$ to obtain fast convergence in calculating reaction rates \cite{Zubarev}.
Then, further Lagrange parameters appear in the generalized Gibbs distribution such as parameters related to the single particle distribution function.
If we consider only single-particle operators in $\rho_{\rm rel}(t')$, in the standard lowest approximations no quantum correlations and bound states are described. This long-standing problem of transport models using kinetic equations can be solved by the NSO method including few-particle correlations into the set of relevant observables \cite{Ropke:1988ymk,Danielewicz:1991dh,Kuhrts:2000zs}.  
It is supposed that the NSO does not depend on the choice of the set of relevant observables $\{B_i\}$ if the limit $\epsilon \to 0$ is correctly performed.
An exception are conserved quantities which must be included in the set of relevant observables.
All other correlations in the system are produced dynamically.

In this work, we are dealing with the elimination of the Lagrange parameters $\lambda_n(t'), \lambda_p(t')$. This is the nonequilibrium version of the EOS $n_\tau(T, \mu_{\tau'})$ which relates the chemical potentials $\mu_n,\mu_p$ to the densities $n_n,n_p$.
Because the generalized Gibbs distribution is of exponential form, quantum statistical methods can be used to calculate correlation functions, in particular the single-particle spectral function. 
We perform a cluster expansion of the self-energy for the interacting system \cite{Ropke:2014fia}, and partial densities of different clusters $\{A,Z\}$ are introduced. 
The relevant yields $Y^{\rm rel}_{AZ}$ are calculated as 
\begin{equation} 
\label{Y0}
Y^{\rm rel}_{AZ} \propto R_{AZ}(\lambda_T) \left(\frac{2 \pi \hbar^2}{A m \lambda_T}\right)^{-3/2}
e^{(B_{AZ}+(A-Z) \lambda_n+Z  \lambda_p)/ \lambda_T},
\end{equation}  
where $B_{AZ}$ denotes the (ground state) binding energy and $g_{AZ}$ the degeneracy, $m$ is the nucleon mass.
The prefactor $R_{AZ}(\lambda_T)=g_{AZ}+\sum^{\rm exc}_i g_{AZi} e^{-E_{AZi}/\lambda_T}$ is related to the intrinsic partition function of the cluster $\{A,Z\}$.
The summation is performed over all excited states, excitation energy $E_{AZi}$ and degeneracy $g_{AZi}$.
To obtain the virial form, the continuum contributions have to be included.
For instance, the Beth-Uhlenbeck formula expresses the contribution of the continuum to the intrinsic partition function via the scattering phase shifts, see \cite{Schmidt:1990oyr}. 
The simple nuclear statistical equilibrium (NSE) distribution is obtained for $R_{AZ}=g_{AZ}$, i.e., neglecting the contribution of all excited states and continuum correlations.

We have outlined the connection to the theory of nonequilibrium statistics in order to show that the use of a (local) thermodynamic equilibrium is not misleading but is, in the formulation with a relevant statistical operator, an important ingredient to describe the nonequilibrium process.
The correct implementation of the quasi-equilibrium distribution as relevant statistical operator into the nonequilibrium statistical operator (\ref{rhoZ}) enables us to treat correlations in a systematic way, for instance using quantum statistical methods and diagram techniques. 
A particular issue is the few-particle in-medium Schroedinger equation \cite{Ropke:1983lbc} which contains the self-energy shifts of the nucleons as well as the Pauli blocking for the interaction.
From this, a generalized Beth-Uhlenbeck formula can be derived \cite{Schmidt:1990oyr}. 
A consequence is the lowering of the binding energy $B_{AZ}(T,n)$ with increasing density.
Due to Pauli blocking, bound states are dissolved at a critical value of density that we denote as Mott density $n_{A,Z}^{\rm Mott}(T,P)$ \cite{Ropke:1983lbc}, which depends on $T$ but also on the c.m. momentum $P$ of the cluster.

\section{Freeze-out temperature as function of the baryonic chemical potential} 

Simple empirical approaches such as the NSE have been used to infer a temperature $T_{\rm cf}$ and a chemical potential $\mu_{\rm cf}$ (we fix the total proton fraction $Y_p=n_p/(n_p+n_n)$ so that the baryonic chemical potential $\mu=(1-Y_p) \mu_n+Y_p\mu_p$ is sufficient). 
Already in this simple approximation, it is an amazing fact that particle production from heavy-ion collisions can be interpreted by freeze-out models which consider a nuclear statistical equilibrium with only these two parameters\footnote{The volume $V$, that is generally needed, is typically determined by the most abundant particle or eliminated when ratios of particle yields are fitted. At the highest energies (LHC) for instance the most abundant particles are pions.}. 
Within a more sophisticated approach the feed-down and in-medium effects have to be taken into account.
This improves the reproduction of the observed yields within the freeze-out approach as shown, for instance, for the LHC range \cite{Andronic:2017pug,Donigus:2022haq}, for HIC in the Fermi-energy range \cite{Qin:2011qp,Pais:2019jstw}, or at fission \cite{Ropke:2020hbm}.
Nevertheless, it is amazing that the global form of the distribution of observed yields is already nicely described within simple model calculations as employed in this Section. 

Here we consider the fitted freeze-out parameters derived from the observed yields, as found in the literature. 
Depending on the collision energy ($s_{NN}$), values for the pairs $T_{\rm cf}, \mu_{\rm cf}$ are inferred that define a region in the $T-\mu$ plane.

The beam energy scan (BES) programs of HIC provide insights into the systematics of particle production under varying conditions of temperature and density of the evolving hadronic fireball created in these experiments. 
The yields of hadrons and nuclei are analyzed using the thermal statistical model of hadron production.
It is a remarkable fact that the statistical model makes excellent reproduction of particle yields with just two free parameters: the temperature $T_{\rm cf}$ and the baryon chemical potential $\mu_{\rm cf}$ if we neglect the dependence on the asymmetry parameter $Y_p$
\cite{Andronic:2017pug}, which is the proton fraction. In the thermal model fits it is usually determined through the charge-to-baryon-number ratio ($Q/B$) of the initially collided nuclei. 
{As discussed below, the dependence on $Y_p$ near $Y_p=0.5$ is weak.}

A statistical model analysis of various experiments at different facilities (LBNL, GSI LINAC \& SIS, BNL AGS, CERN SPS, BNL RHIC, CERN LHC) and different collision energies (few \unit{\MeV} to \SI{5}{\TeV}), has been performed in a series of works.
A prominent one being \cite{Cleymans:2005xv}, followed by similar analyses incorporating the ever-growing set of experimental data from facilities running experiments such as HADES, E871, NA49/NA61, STAR, ALICE and others.

\begin{figure}[!ht] 
\begin{center}
 	\includegraphics[width=0.8\textwidth]{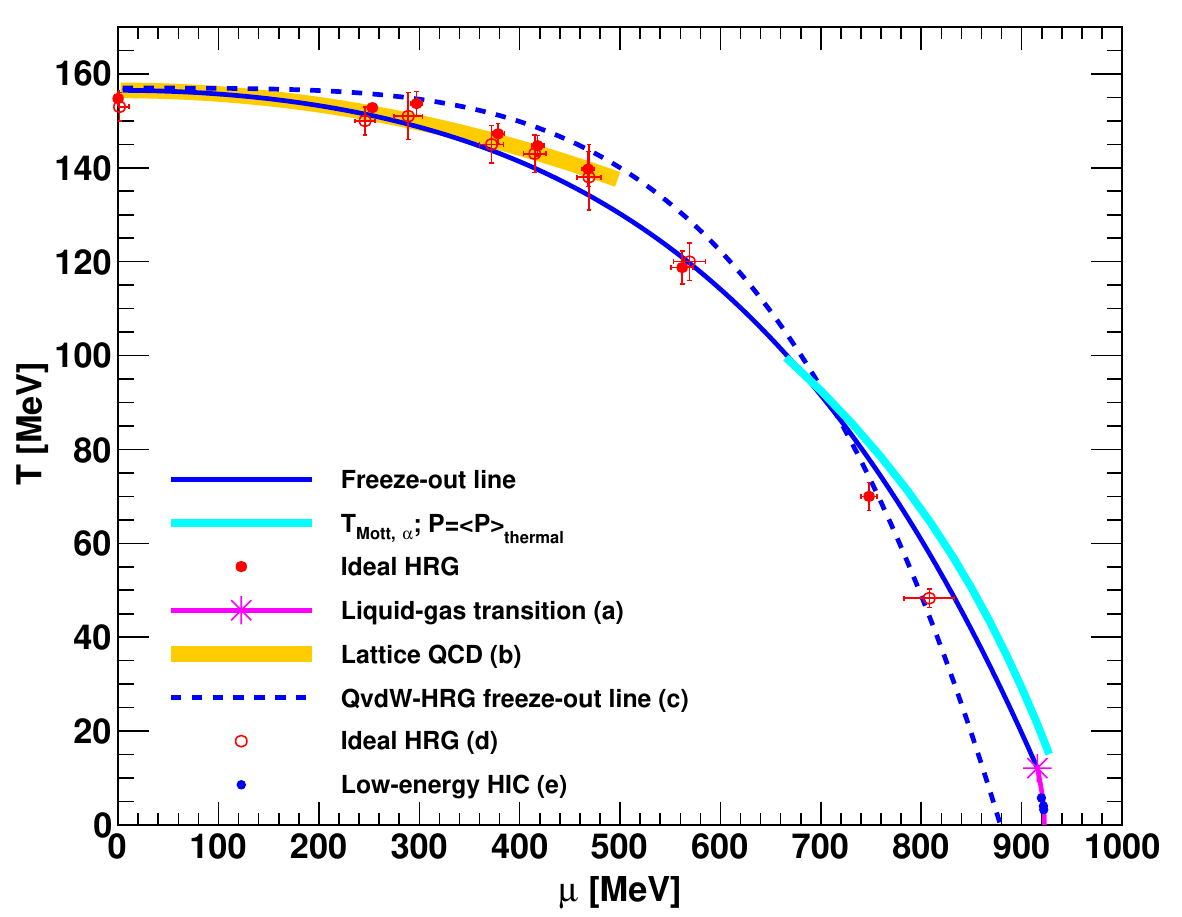}
	\caption{Chemical freeze-out temperature $T_{\rm cf}$ as function of the baryonic chemical potential $\mu$. The parametrization from Eq.~(\ref{Eq:fit}) is depicted as solid blue line. The liquid-gas nuclear matter phase transition is also shown (magenta line) ending at the critical point (magenta star), obtained from a quantum statistical approach \cite{Typel:2009sy} (a).
 The yellow lattice QCD band (b) is from Ref. \cite{HotQCD:2019xnw}. The red data points shown are from ideal hadron resonance gas (HRG) fits to measurements at different collision energies (ALICE~\cite{ALICE:2013mez,ALICE:2013cdo,ALICE:2013xmt,ALICE:2014jbq}, NA49~\cite{NA49:2002pzu,NA49:2006gaj,NA49:2008ysv,NA49:2008goy,NA49:2004irs,NA49:2011blr,Friese:2002re,NA49:2011bfu,NA49:2007stj}, E802/E866~\cite{E-802:1999acy,E802:1999hit,E-802:1999ewk,Becattini:2000jw}, HADES~\cite{HADES:2015oef}, GSI/SIS~\cite{Cleymans:1998yb,Averbeck:2000sn}), see the text for references. Additionally, the fits from Ref.~\cite{Poberezhnyuk:2019pxs} are shown in open symbols (d), where also a parametrization is given for the freeze-out curve if the quantum van der Waals hadron resonance gas is used~\cite{Poberezhnyuk:2019pxs} (c), shown as dashed blue line. Furthermore, results from measurements of the low-energy heavy-ion collisions are shown~\cite{Natowitz:2010ti} (e). The cyan line shows the Mott line for $\alpha$ clusters at thermally averaged momentum, see Fig.~\ref{Fig:2} and Fig.~\ref{Fig:3}.}
 \label{Fig:1} 
 \end{center}
 \end{figure}

Experiments at highest energies \cite{Andronic:2017pug} from LHC give $T_{\rm cf}=\,$\SI[separate-uncertainty=true]{156.5 \pm 1.5}{\MeV}
which coincides with the pseudocritical temperature of the chiral crossover $T_{c}=\,$\SI[separate-uncertainty=true]{156.5 \pm 1.5}{\MeV} obtained from lattice QCD simulations at zero baryon chemical potential $\mu=$\,\num{0} \cite{HotQCD:2018pds}.
Other experiments at lower energies give lower temperatures at increasing chemical potentials, in accordance with lattice simulations up to $\mu \approx$\,\SI{300}{\MeV} \cite{Borsanyi:2020fev}.
Values down to $T_{\rm cf}=$\,\SI{48.3}{\MeV} are obtained from data for HIC at $E_{\rm lab}=$\,\SI{0.8}{A \GeV}, also shown in Fig. \ref{Fig:1}. 
On the other hand, HIC experiments at low laboratory energies of $E_{\rm lab} =$\,\SI{35}{A \MeV}\,\cite{Kowalski:2006ju} were analysed in the quantum statistical freeze-out scheme in 
\cite{Natowitz:2010ti,Qin:2011qp,Hagel:2014wja}, and freeze-out temperatures in the range \SI{4}{\MeV} $< T_{\rm cf} <$ \SI{10}{\MeV} were reported.
A similar experiment has been performed at GANIL~\cite{Pais:2019jstw}. The corresponding chemical potentials are also shown in Fig.~\ref{Fig:1}.
These parameter values are related to the nuclear matter phase transition, starting with $\mu=$\,\SI{922.8}{\MeV} at  $T=$\,\num{0} and ending with the critical point at $T_{\rm liquid-gas}$ = \SI{12.1}{\MeV}, $\mu_{\rm liquid-gas}$ = \SI{915.61}{\MeV}~\cite{Typel:2009sy}.

In Fig.~\ref{Fig:1}, for the freeze-out temperature $T_{\rm cf}$ as function of the chemical potential the relation

\begin{equation}
\label{Eq:fit}
 T_{\rm cf}(\mu)\,[{\rm MeV}] =  156.5 - 76.68~(\mu\,[{\rm GeV}])^2-139.7~(\mu \,[{\rm GeV}])^4  
\end{equation} 
is shown which interpolates between both limits. 
The first two coefficients are determined by the lattice calculations, whereas the third coefficient is chosen such that the critical point for the nuclear matter phase transition is met.
This interpolation formula is similar to the relation proposed by Cleymans {\it et al.}~\cite{Cleymans:2005xv}.
A more recent parametrization was obtained by Vovchenko et al.~\cite{Vovchenko:2015idt} including new data. A slightly different expression was given recently in Ref. \cite{Poberezhnyuk:2019pxs}
which is obtained within a quantum van der Waals hadron resonance gas (QvdW-HRG) model.

In this contribution, we would like to focus on the region between these two extremes, where the transition from baryon stopping to nuclear transparency takes place and the highest baryon densities at freeze-out are reached. 
It is the region of c.m.s. energies of the future NICA facility \cite{Kekelidze:2016hhw}, $\sqrt{s_{NN}}=$\,\SIrange{2}{11}{\GeV}, which was partly addressed already by AGS and CERN-SPS experiments as well as the RHIC BES I program and the recent HADES experiment at GSI. 
Besides the MPD and BM@N experiments at NICA it will be covered in future by the low-energy RHIC and the RHIC fixed target programs as well as the FAIR CBM experiment.
It has been suggested in \cite{Ropke:2017dur} that both regions, highest and lowest $T$, should be joined, and that new experiments can help to investigate the entire temperature range.

 We introduced the relation Eq.\,(\ref{Eq:fit}) as an interpolation between high and low $T$ regions.
 We see that experimental data fit nicely this plot.
 In the following Section we discuss the cluster formation in matter which could explain the course of the freeze-out line.  \\
 
\section{Freeze-out in the density-temperature diagram and Mott lines}

For given $T,\mu,Y_p$, the calculation of the baryon number density (EoS) and the composition of interacting nucleonic matter is an intricate problem.
We need a microscopic approach to understand the processes of cluster formation.
In particular, the in-medium effects are determined by densities and the composition which must be calculated self-consistently.

When analysing the particle yields from HIC, often a simple model of nuclear statistical equilibrium (NSE) is used to extract the freeze-out values $T_{\rm cf}, \mu_{\rm cf}$. It became evident that this is not sufficient because the densities are high so that medium effects have to be considered \cite{Qin:2011qp}. 
Self-energy effects are treated in different approximations, one of them is the RMF-DD2 approximation used in our calculations \cite{Typel:2009sy}.
Crucial for the yields of light clusters is the Pauli blocking which leads to the suppression of cluster formation at high densities \cite{Ropke:1983lbc}. 
Quantum statistical calculations were performed \cite{Ropke:2017dur}, but simple approximations are given by the model of excluded volume \cite{Hempel:2009mc}, for a detailed description see Ref. \cite{Typel:2009sy}.

Considering the freeze-out line in the $T-\mu$ plane, we have three different regions: 
(i) nuclear matter at temperatures up to \SI{40}{\MeV} where nuclear clusters contribute about \SI{50}{\%} to the baryon density and resonances are of minor importance, (ii) the region \SIrange{40}{130}{\MeV} where clusters dissociate and resonances are of relevance, and (iii) above the temperature \SI{130}{\MeV} of the maximal baryon density at freeze-out, where we have pion-dominated matter with negligible net baryon density.

\begin{figure}[!ht] 
\begin{center}
 \includegraphics[width=.8\textwidth]{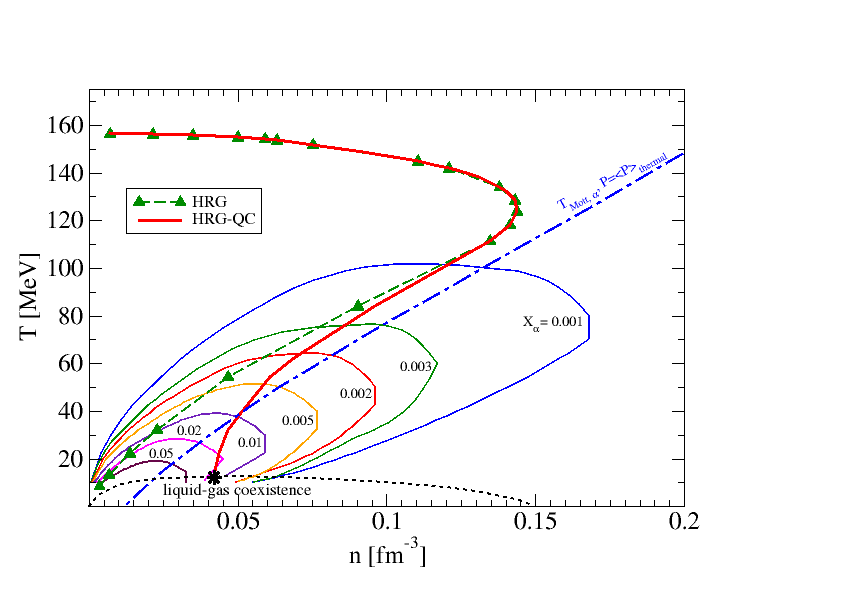}
	\caption{Chemical freeze out lines for the hadron resonance gas (HRG) without and with (HRG-QC) quantum correlations in the temperature density plane (phase diagram) for the relation (\ref{Eq:fit}).
  The coexistence region for the nuclear liquid-gas transition (black solid line) 
 is shown together with its critical endpoint (asterisk). 
 Lines of constant $\alpha$-particle mass fraction $X_\alpha$ are shown as labelled solid lines. 
 The maxima of the $\alpha$ fraction at given temperatures $X^{\rm max}_\alpha(T)$ lie close to the Mott line for $\alpha$ clusters at thermally averaged momentum (dash-dotted line), see also Fig. \ref{Fig:3}.
 The asymmetry $Y_p=$\,\num{0.4} is taken.
 }
 \label{Fig:2} 
 \end{center}
 \end{figure}

We start with the ideal hadron resonance gas which considers in addition to the nucleons $n,p$ also all mesons up to the mass of $K_4^*(2045)$ and baryons up to $\Omega^-$ as they are listed, together with their degeneracy, in the Review of Particle Physics~\cite{ParticleDataGroup:2020ssz}, according to the choice of Albright, Kapusta and Young\,\cite{Albright:2014gva} which has also been made by Begun, Gazdzicki and Gorenstein \cite{Begun:2015ifa}.
We used the hybrid EOS from Ref.\,\cite{Mir:2023wkm,Kahangirwe:2024xyl} and calculated the baryon number density (baryon density minus antibaryon density) for freeze-out conditions, Eq. (\ref{Eq:fit}).
This calculation takes excluded volume effects into account.
The result for $n(T,\mu)$ for $T=T_{\rm cf}(\mu)$ according to Eq. (\ref{Eq:fit}) is shown in Fig. \ref{Fig:2} (green dashed line)\footnote{We tested the dependence on the $Q/B$ ratio (\num{0.4} vs. \num{0.5}) and see no strong dependence, see Fig.~\ref{Fig:3} below. }.
A baryon radius of $r_B =$\SI{0.365}{fm} according to 
 Vitiuk {\it et al.} \cite{Vitiuk:2020jsa} is chosen to determine the excluded volume. We obtain a maximum freeze-out density \SI{0.144}{\per\cubic\femto\metre}. Similar values for the maximum freeze-out density were discussed in Refs. \cite{Randrup:2006nr,Poberezhnyuk:2019pxs}.
The ideal hadron resonance gas is not appropriate at high densities because it neglects in-medium effects. (For the ideal HRG, a maximum freeze-out density \SI{0.175}{\per\cubic\femto\metre} is obtained at $T=$\SI{130}{\MeV}, to be compared with the maximum value \SI{0.225}{\per\cubic\femto\metre} given in Ref.~\cite{Poberezhnyuk:2019pxs}.)

Our new result is the density $n(T,\mu)$ for $T_{\rm cf}(\mu)$ (HRG-QC, red line) which takes quantum correlations within the nucleon subsystem into account.
To this end, we use a quantum statistical approach to the neutron-proton system \cite{Typel:2009sy} which contains in addition to the self-energies, calculated in RMF-DD2 approximation, also the formation of light nuclei ($^2$H, $^3$H, $^3$He, $^4$He). 
We calculate the in-medium shifts of the binding energies, in particular the Pauli blocking shifts, so that the bound states disappear with increasing density due to the so-called Mott effect. 
All components of the non-ideal HRG except neutrons and protons are treated according to the approach described above. These contributions are added to the 
results of the quantum statistical approach to the neutron-proton system. 
Above $T=90$ MeV the density of clusters becomes very low, and the non-ideal HRG remains as a good approximation.

However, there is a large discrepancy between the HRG and the HRG-QC result in Fig.\,\ref{Fig:2}, in particular at temperatures below 80 MeV when cluster formation is not taken into account. 
For instance, in \cite{Poberezhnyuk:2019pxs} it was found that at the lowest considered energy of $\sqrt{s_{NN}} =$\,\SI{1.9}{\GeV} freeze-out in both models, the ideal and the quantum van der Waals hadron resonance gas, takes place at baryon density of $n =$\,\SI{2.17d-4}{\per\cubic\femto\metre}. 
HIC experiments performed at low energies \cite{Qin:2011qp} obtained fragments at temperatures of about \SI{10}{\MeV} and values for the freeze-out density in the range of \SI{0.03}{\per\cubic\femto\metre}. 
The formation of correlations and bound states has to be treated correctly in the region of low temperatures and high baryon densities, using a quantum statistical approach. 

 Calculating the contributions of light clusters to the density $n(T_{\rm cf},\mu)$,  
 an important in-medium effect is Pauli blocking which gives a reduction of the binding energy. 
 This shift depends on $T, n_p, n_n$, but also on the c.m. momentum $P$.
 Values and parametrisations are found in \cite{Ropke:2014fia}, see also \cite{Typel:2009sy}. 
 In particular it determines the so-called Mott density where the bound state is dissolved. 
 To show the region in the phase diagram where bound states are formed, 
we focus on the $\alpha$ particles which are well bound clusters.
Contour lines for the mass fraction $X_\alpha =4 n_\alpha/n$ of $\alpha$ particles which depend on $T,n,Y_p$ are shown in Fig. \ref{Fig:2}. 

In logarithmic representation, Fig. \ref{Fig:3}, the banana-like region of $\alpha$ cluster formation is shown in comparison with the Mott line which describes the site in the phase diagram where the bound state disappears. Such diagrams were first presented in \cite{Ropke:1983lbc} ($\alpha$-particle association degree), and a Mott line was shown which characterizes the dissolution of the $\alpha$-particles due to Pauli blocking. Since the Pauli blocking shift strongly depends on the c.m. momentum $P$, we show the Mott line for the dissolution of $\alpha$-particles at thermally averaged momentum-squared $P^2=(3/2)m_\alpha T$~\cite{Schulz:1983iho,Ropke:2011tr}, with $m_\alpha = 4 m_N$. It is close to the line $X_\alpha^{\rm max}(T)$ where at given $T$ the mass fraction of $\alpha$ particles takes its maximum value. 

The Mott line for the dissolution of $\alpha$-particles at thermally averaged momentum is also shown in Fig. \ref{Fig:2}. 
In the temperature range from \SIrange{30}{130}{\MeV} it can be well described by a linear fit formula
\begin{equation}
\label{eq:fitTn}
    T_{\rm Mott, \alpha}(n){\rm [MeV]} = 2.6 + 731.9 ~ n{\rm [fm^{-3}]}~.
\end{equation}

The close neighborhood of the Mott line $T_{\rm Mott, \alpha}(n)$ with the chemical freeze-out line (HRG-QC) below $T \approx$\,\SI{130}{\MeV} is an interesting result of our work. It indicates that the chemical freeze-out process in expanding matter may be related to the formation of bound states. The Mott effect which describes the ability to form bound states can be considered as a strong change of the intrinsic structure of nuclear matter where the relaxation to chemical equilibrium is suppressed.
We suggest to consider the bound state formation at the Mott line as a prerequisite for chemical freeze-out since it entails a sudden change in the microscopic state of the system.

\begin{figure}[!ht] 
\begin{center}
    \includegraphics[width=.8\textwidth]{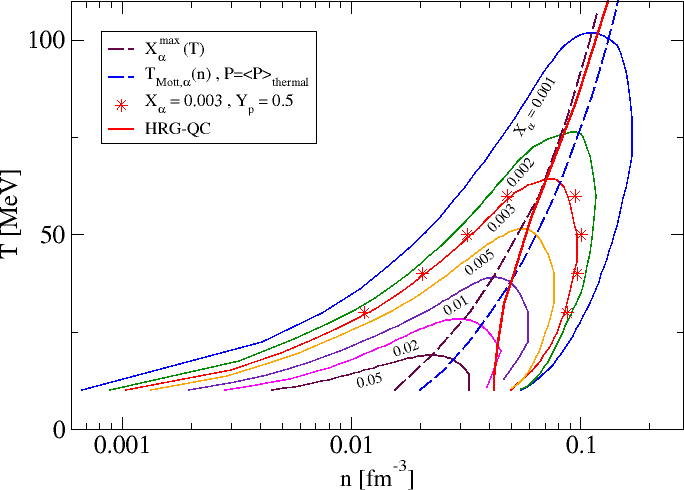}
	\caption{ Contour lines for mass fractions of $\alpha$ clusters (logarithmic baryon density scale) at proton fraction $Y_p =$\,\num{0.4}. In addition, the Mott line at thermally averaged momentum (see text) and mass fraction maximum $X^{\rm max}_\alpha(T)$ at given temperature line are shown. The red stars indicate a calculation for the case of symmetric matter, $Y_p =$\,\num{0.5}, and a mass fraction of 0.003.}
 \label{Fig:3} 
 \end{center}
 \end{figure}

To show the dependence of the asymmetry $Y_p$, in Fig. \ref{Fig:3} results for the asymmetry 
$Y_p =$\,\num{0.5} (star, for $X_\alpha =$\,\num{0.003}) are presented. 
The effect of asymmetry is small.
Below $T=$\,\SI{10}{\MeV} the calculation with only four species of light nuclei is not sufficient because larger nuclei will occur. In addition, the region of the liquid-gas phase transition is reached there.

\section{Conclusions} 
We justified the use of quasi-equilibrium information to describe the nonequilibrium expansion of the hot fireball produced in HIC. 
The freeze-out concept is related to fast processes which change the chemical composition of the hot and dense matter. 
We propose that the formation of bound states with decreasing density and/or temperature can be considered as the microscopic process responsible for the freeze-out phenomenon. The freeze-out parameters for this formation process coincide with those of the reverse process describing the bound state dissociation (Mott effect).

Our aim is to connect different asymptotic regions where the freeze-out concepts are used to explain the observed yields: heavy-ion collisions at ultrarelativistic energies \cite{Andronic:2017pug} and collision \cite{Qin:2011qp} as well as fission \cite{Ropke:2020hbm} processes at low energies. 
After a first attempt for a unified description \cite{Ropke:2017dur}, a more detailed approach was given in \cite{Poberezhnyuk:2019pxs} but without describing nucleonic correlations and cluster formation in the low-temperature region of hadronic matter. 
We present a quantum statistical approach which describes formation of bound states and their dissolution due to Pauli blocking.

Simple statistical models used to extract the freeze-out parameters are working well in the limiting cases of high temperatures \cite{Andronic:2017pug,Donigus:2022haq} and in the low-temperature, dilute matter region
\cite{Qin:2011qp,Ropke:2020hbm}.
In the intermediate range of temperatures and densities, the application of simple statistical models is problematic because in-medium effects become important. The method of the nonequilibrium statistical operator allows a systematic inclusion of many-particle effects.

The use of the freeze-out concept remains an important ingredient to understand the observed yields of particles produced in the expanding hot and dense matter of an ultrarelativistic heavy-ion collision.
We propose in this work a new chemical freeze-out line in the temperature-density plane, connecting low and high temperature regime. 
We find that this new freeze-out line in the intermediate region of temperatures 
\num{20}\,$\lesssim T$\,[\unit{\MeV}]\,$\lesssim$ \,\num{120} is well-correlated with the maximum of the $\alpha$ cluster fraction at a given temperature.
We also show that the Mott-line for $\alpha$ particles evaluated at their thermally averaged momentum is a good approximation for the chemical freeze-out line in this intermediate temperature-density range. 
While this chemical freeze-out line is well defined in the $T-\mu$ plane, see Eq. (\ref{Eq:fit}), its trajectory in the temperature-density diagram is sensitive to the treatment of interparticle correlations like Pauli blocking or excluded volume effects. Our result Eq. (\ref{eq:fitTn}), which represents the  thermally averaged Mott-line for $\alpha$ particles in the range $30 \le T {\rm [MeV]} \le 130$, can be considered as a good approximation for the chemical freeze-out line in the temperature-density diagram.

We expect that future experiments like NICA, FAIR, CERN-SPS, RHIC-BES and others will provide new data 
on particle production in the intermediate temperature region we have considered in this work. The results of the present work have shown that medium effects on correlations will play an important role in the interpretation of these experiments within a freeze-out picture.
We conclude that medium effects, being often small corrections, can be essential in the intermediate temperature range.

\subsection*{Acknowledgments}
D.B. was supported by NCN under grant No. 2021/43/P/ST2/03319. G.R. acknowledges a honorary stipend from the Foundation for Polish Science within the Alexander von Humboldt programme under grant No. DPN/JJL/402-4773/2022. B.D. acknowledges the support from Bundesministerium f\"{u}r Bildung und Forschung through ErUM-FSP T01 (F\"{o}rderkennzeichen 05P21RFCA1).



\begin{thebibliography}{10}
\expandafter\ifx\csname url\endcsname\relax
  \def\url#1{\texttt{#1}}\fi
\expandafter\ifx\csname urlprefix\endcsname\relax\def\urlprefix{URL }\fi
\expandafter\ifx\csname href\endcsname\relax
  \def\href#1#2{#2} \def\path#1{#1}\fi

\bibitem{Gutbrod:1976zzr}
H.~H. Gutbrod, A.~Sandoval, P.~J. Johansen, A.~M. Poskanzer, J.~Gosset, W.~G.
  Meyer, G.~D. Westfall, R.~Stock, {Final State Interactions in the Production
  of Hydrogen and Helium Isotopes by Relativistic Heavy Ions on Uranium}, Phys.
  Rev. Lett. 37 (1976) 667--670.
\newblock \href {https://doi.org/10.1103/PhysRevLett.37.667}
  {\path{doi:10.1103/PhysRevLett.37.667}}.

\bibitem{Gosset:1976cy}
J.~Gosset, H.~H. Gutbrod, W.~G. Meyer, A.~M. Poskanzer, A.~Sandoval, R.~Stock,
  G.~D. Westfall, {Central Collisions of Relativistic Heavy Ions}, Phys. Rev. C
  16 (1977) 629--657.
\newblock \href {https://doi.org/10.1103/PhysRevC.16.629}
  {\path{doi:10.1103/PhysRevC.16.629}}.

\bibitem{Westfall:1976fu}
G.~D. Westfall, J.~Gosset, P.~J. Johansen, A.~M. Poskanzer, W.~G. Meyer, H.~H.
  Gutbrod, A.~Sandoval, R.~Stock, {Nuclear Fireball Model for Proton Inclusive
  Spectra from Relativistic Heavy Ion Collisions}, Phys. Rev. Lett. 37 (1976)
  1202--1205.
\newblock \href {https://doi.org/10.1103/PhysRevLett.37.1202}
  {\path{doi:10.1103/PhysRevLett.37.1202}}.

\bibitem{Andronic:2017pug}
A.~Andronic, P.~Braun-Munzinger, K.~Redlich, J.~Stachel, {Decoding the phase
  structure of QCD via particle production at high energy}, Nature 561~(7723)
  (2018) 321--330.
\newblock \href {http://arxiv.org/abs/1710.09425} {\path{arXiv:1710.09425}},
  \href {https://doi.org/10.1038/s41586-018-0491-6}
  {\path{doi:10.1038/s41586-018-0491-6}}.

\bibitem{Vitiuk:2020jsa}
O.~V. Vitiuk, K.~A. Bugaev, E.~S. Zherebtsova, D.~B. Blaschke, L.~V. Bravina,
  E.~E. Zabrodin, G.~M. Zinovjev, {Resolving the hyper-triton yield description
  puzzle in high energy nuclear collisions}, Eur. Phys. J. A 57~(2) (2021) 74.
\newblock \href {http://arxiv.org/abs/2007.07376} {\path{arXiv:2007.07376}},
  \href {https://doi.org/10.1140/epja/s10050-021-00370-6}
  {\path{doi:10.1140/epja/s10050-021-00370-6}}.

\bibitem{Bugaev:2020sgz}
K.~A. Bugaev, et~al., {Second virial coefficients of light nuclear clusters and
  their chemical freeze-out in nuclear collisions}, Eur. Phys. J. A 56~(11)
  (2020) 293.
\newblock \href {http://arxiv.org/abs/2005.01555} {\path{arXiv:2005.01555}},
  \href {https://doi.org/10.1140/epja/s10050-020-00296-5}
  {\path{doi:10.1140/epja/s10050-020-00296-5}}.

\bibitem{Cleymans:2005xv}
J.~Cleymans, H.~Oeschler, K.~Redlich, S.~Wheaton, {Comparison of chemical
  freeze-out criteria in heavy-ion collisions}, Phys. Rev. C 73 (2006) 034905.
\newblock \href {http://arxiv.org/abs/hep-ph/0511094}
  {\path{arXiv:hep-ph/0511094}}, \href
  {https://doi.org/10.1103/PhysRevC.73.034905}
  {\path{doi:10.1103/PhysRevC.73.034905}}.

\bibitem{Vovchenko:2015idt}
V.~Vovchenko, V.~V. Begun, M.~I. Gorenstein, {Hadron multiplicities and
  chemical freeze-out conditions in proton-proton and nucleus-nucleus
  collisions}, Phys. Rev. C 93~(6) (2016) 064906.
\newblock \href {http://arxiv.org/abs/1512.08025} {\path{arXiv:1512.08025}},
  \href {https://doi.org/10.1103/PhysRevC.93.064906}
  {\path{doi:10.1103/PhysRevC.93.064906}}.

\bibitem{Poberezhnyuk:2019pxs}
R.~Poberezhnyuk, V.~Vovchenko, A.~Motornenko, M.~I. Gorenstein, H.~Stoecker,
  {Chemical freeze-out conditions and fluctuations of conserved charges in
  heavy-ion collisions within quantum van der Waals model}, Phys. Rev. C
  100~(5) (2019) 054904.
\newblock \href {http://arxiv.org/abs/1906.01954} {\path{arXiv:1906.01954}},
  \href {https://doi.org/10.1103/PhysRevC.100.054904}
  {\path{doi:10.1103/PhysRevC.100.054904}}.

\bibitem{Heinz:2007in}
U.~W. Heinz, G.~Kestin, {Jozso's Legacy: Chemical and Kinetic Freeze-out in
  Heavy-Ion Collisions}, Eur. Phys. J. ST 155 (2008) 75--87.
\newblock \href {http://arxiv.org/abs/0709.3366} {\path{arXiv:0709.3366}},
  \href {https://doi.org/10.1140/epjst/e2008-00591-4}
  {\path{doi:10.1140/epjst/e2008-00591-4}}.

\bibitem{Blaschke:2011hm}
D.~Blaschke, J.~Berdermann, J.~Cleymans, K.~Redlich, {Chiral condensate and
  Mott-Anderson freeze-out}, Few Body Syst. 53 (2012) 99--109.
\newblock \href {http://arxiv.org/abs/1109.5391} {\path{arXiv:1109.5391}},
  \href {https://doi.org/10.1007/s00601-011-0261-6}
  {\path{doi:10.1007/s00601-011-0261-6}}.

\bibitem{Blaschke:2011ry}
D.~B. Blaschke, J.~Berdermann, J.~Cleymans, K.~Redlich, {Chiral condensate and
  chemical freeze-out}, Phys. Part. Nucl. Lett. 8 (2011) 811--817.
\newblock \href {http://arxiv.org/abs/1102.2908} {\path{arXiv:1102.2908}},
  \href {https://doi.org/10.1134/S154747711108005X}
  {\path{doi:10.1134/S154747711108005X}}.

\bibitem{Povh:1987ju}
B.~Povh, J.~H{\"u}fner, {Geometric Interpretation of Hadron Proton Total
  Cross-sections and a Determination of Hadronic Radii}, Phys. Rev. Lett. 58
  (1987) 1612--1615.
\newblock \href {https://doi.org/10.1103/PhysRevLett.58.1612}
  {\path{doi:10.1103/PhysRevLett.58.1612}}.

\bibitem{Hippe:1995hu}
H.~J. Hippe, S.~P. Klevansky, {Nambu-Jona-Lasinio model compared with chiral
  perturbation theory: The Pion radius in SU(2) revisited}, Phys. Rev. C 52
  (1995) 2172--2184.
\newblock \href {https://doi.org/10.1103/PhysRevC.52.2172}
  {\path{doi:10.1103/PhysRevC.52.2172}}.

\bibitem{TMEP:2022xjg}
H.~Wolter, et~al., {Transport model comparison studies of intermediate-energy
  heavy-ion collisions}, Prog. Part. Nucl. Phys. 125 (2022) 103962.
\newblock \href {http://arxiv.org/abs/2202.06672} {\path{arXiv:2202.06672}},
  \href {https://doi.org/10.1016/j.ppnp.2022.103962}
  {\path{doi:10.1016/j.ppnp.2022.103962}}.

\bibitem{Glassel:2021rod}
S.~Glä\ss{}el, V.~Kireyeu, V.~Voronyuk, J.~Aichelin, C.~Blume,
  E.~Bratkovskaya, G.~Coci, V.~Kolesnikov, M.~Winn, {Cluster and hypercluster
  production in relativistic heavy-ion collisions within the
  parton-hadron-quantum-molecular-dynamics approach}, Phys. Rev. C 105~(1)
  (2022) 014908.
\newblock \href {http://arxiv.org/abs/2106.14839} {\path{arXiv:2106.14839}},
  \href {https://doi.org/10.1103/PhysRevC.105.014908}
  {\path{doi:10.1103/PhysRevC.105.014908}}.

\bibitem{Kireyeu:2024woo}
V.~Kireyeu, G.~Coci, S.~Gl\"a\ss{}el, J.~Aichelin, C.~Blume, E.~Bratkovskaya,
  {Cluster formation near midrapidity: How the production mechanisms can be
  identified experimentally}, Phys. Rev. C 109~(4) (2024) 044906.
\newblock \href {https://doi.org/10.1103/PhysRevC.109.044906}
  {\path{doi:10.1103/PhysRevC.109.044906}}.

\bibitem{Ropke:1988ymk}
G.~R\"opke, H.~Schulz, {Including cluster formation into a
  Vlasov-Uehling-Uhlenbeck approach to intermediate energy heavy ion
  collisions}, Nucl. Phys. A 477 (1988) 472--486.
\newblock \href {https://doi.org/10.1016/0375-9474(88)90352-1}
  {\path{doi:10.1016/0375-9474(88)90352-1}}.

\bibitem{Danielewicz:1991dh}
P.~Danielewicz, G.~F. Bertsch, {Production of deuterons and pions in a
  transport model of energetic heavy ion reactions}, Nucl. Phys. A 533 (1991)
  712--748.
\newblock \href {https://doi.org/10.1016/0375-9474(91)90541-D}
  {\path{doi:10.1016/0375-9474(91)90541-D}}.

\bibitem{Kuhrts:2000zs}
C.~Kuhrts, M.~Beyer, P.~Danielewicz, G.~R{\"o}pke, {Medium corrections in the
  formation of light charged particles in heavy ion reactions}, Phys. Rev. C 63
  (2001) 034605.
\newblock \href {http://arxiv.org/abs/nucl-th/0009037}
  {\path{arXiv:nucl-th/0009037}}, \href
  {https://doi.org/10.1103/PhysRevC.63.034605}
  {\path{doi:10.1103/PhysRevC.63.034605}}.

\bibitem{Kanada-Enyo:2012yif}
Y.~Kanada-En'yo, M.~Kimura, A.~Ono, {Antisymmetrized molecular dynamics and its
  applications to cluster phenomena}, PTEP 2012 (2012) 01A202.
\newblock \href {http://arxiv.org/abs/1202.1864} {\path{arXiv:1202.1864}},
  \href {https://doi.org/10.1093/ptep/pts001} {\path{doi:10.1093/ptep/pts001}}.

\bibitem{Gross:2022hyw}
F.~Gross, et~al., {50 Years of Quantum Chromodynamics}, Eur. Phys. J. C 83
  (2023) 1125.
\newblock \href {http://arxiv.org/abs/2212.11107} {\path{arXiv:2212.11107}},
  \href {https://doi.org/10.1140/epjc/s10052-023-11949-2}
  {\path{doi:10.1140/epjc/s10052-023-11949-2}}.

\bibitem{Typel:2009sy}
S.~Typel, G.~R{\"o}pke, T.~Kl{\"a}hn, D.~Blaschke, H.~H. Wolter, {Composition
  and thermodynamics of nuclear matter with light clusters}, Phys. Rev. C 81
  (2010) 015803.
\newblock \href {http://arxiv.org/abs/0908.2344} {\path{arXiv:0908.2344}},
  \href {https://doi.org/10.1103/PhysRevC.81.015803}
  {\path{doi:10.1103/PhysRevC.81.015803}}.

\bibitem{Zubarev}
D.~Zubarev, et~al., {Statistical Mechanics of Non-equilibrium Processes},
  Wiley, 1996.

\bibitem{Ropke:2014fia}
G.~R\"opke, {Nuclear matter equation of state including two-, three-, and
  four-nucleon correlations}, Phys. Rev. C 92~(5) (2015) 054001.
\newblock \href {http://arxiv.org/abs/1411.4593} {\path{arXiv:1411.4593}},
  \href {https://doi.org/10.1103/PhysRevC.92.054001}
  {\path{doi:10.1103/PhysRevC.92.054001}}.

\bibitem{Schmidt:1990oyr}
M.~Schmidt, G.~R\"opke, H.~Schulz, {Generalized Beth-Uhlenbeck approach for hot
  nuclear matter}, Annals Phys. 202~(1) (1990) 57--99.
\newblock \href {https://doi.org/10.1016/0003-4916(90)90340-T}
  {\path{doi:10.1016/0003-4916(90)90340-T}}.

\bibitem{Ropke:1983lbc}
G.~R\"opke, M.~Schmidt, L.~M\"unchow, H.~Schulz, {Particle clustering and Mott
  transition in nuclear matter at finite temperature (II)}, Nucl. Phys. A 399
  (1983) 587--602.
\newblock \href {https://doi.org/10.1016/0375-9474(83)90265-8}
  {\path{doi:10.1016/0375-9474(83)90265-8}}.

\bibitem{Donigus:2022haq}
B.~D\"onigus, G.~R\"opke, D.~Blaschke, {Deuteron yields from heavy-ion
  collisions at energies available at the CERN Large Hadron Collider: Continuum
  correlations and in-medium effects}, Phys. Rev. C 106~(4) (2022) 044908.
\newblock \href {http://arxiv.org/abs/2206.10376} {\path{arXiv:2206.10376}},
  \href {https://doi.org/10.1103/PhysRevC.106.044908}
  {\path{doi:10.1103/PhysRevC.106.044908}}.

\bibitem{Qin:2011qp}
L.~Qin, et~al., {Laboratory Tests of Low Density Astrophysical Equations of
  State}, Phys. Rev. Lett. 108 (2012) 172701.
\newblock \href {http://arxiv.org/abs/1110.3345} {\path{arXiv:1110.3345}},
  \href {https://doi.org/10.1103/PhysRevLett.108.172701}
  {\path{doi:10.1103/PhysRevLett.108.172701}}.

\bibitem{Pais:2019jstw}
H.~Pais, et~al., {Low-density in-medium effects on light clusters from
  heavy-ion data}, Phys. Rev. Lett. 125~(1) (2020) 012701.
\newblock \href {http://arxiv.org/abs/1911.10849} {\path{arXiv:1911.10849}},
  \href {https://doi.org/10.1103/PhysRevLett.125.012701}
  {\path{doi:10.1103/PhysRevLett.125.012701}}.

\bibitem{Ropke:2020hbm}
G.~R\"opke, J.~B. Natowitz, H.~Pais, {Nonequilibrium information entropy
  approach to ternary fission of actinides}, Phys. Rev. C 103~(6) (2021)
  061601.
\newblock \href {http://arxiv.org/abs/2012.14691} {\path{arXiv:2012.14691}},
  \href {https://doi.org/10.1103/PhysRevC.103.L061601}
  {\path{doi:10.1103/PhysRevC.103.L061601}}.

\bibitem{HotQCD:2018pds}
A.~Bazavov, et~al., {Chiral crossover in QCD at zero and non-zero chemical
  potentials}, Phys. Lett. B 795 (2019) 15.
\newblock \href {http://arxiv.org/abs/1812.08235} {\path{arXiv:1812.08235}},
  \href {https://doi.org/10.1016/j.physletb.2019.05.013}
  {\path{doi:10.1016/j.physletb.2019.05.013}}.

\bibitem{Borsanyi:2020fev}
S.~Borsanyi, Z.~Fodor, J.~N. Guenther, R.~Kara, S.~D. Katz, P.~Parotto,
  A.~Pasztor, C.~Ratti, K.~K. Szabo, {QCD Crossover at Finite Chemical
  Potential from Lattice Simulations}, Phys. Rev. Lett. 125~(5) (2020) 052001.
\newblock \href {http://arxiv.org/abs/2002.02821} {\path{arXiv:2002.02821}},
  \href {https://doi.org/10.1103/PhysRevLett.125.052001}
  {\path{doi:10.1103/PhysRevLett.125.052001}}.

\bibitem{Kowalski:2006ju}
S.~Kowalski, et~al., {Experimental determination of the symmetry energy of a
  low density nuclear gas}, Phys. Rev. C 75 (2007) 014601.
\newblock \href {http://arxiv.org/abs/nucl-ex/0602023}
  {\path{arXiv:nucl-ex/0602023}}, \href
  {https://doi.org/10.1103/PhysRevC.75.014601}
  {\path{doi:10.1103/PhysRevC.75.014601}}.

\bibitem{Natowitz:2010ti}
J.~B. Natowitz, et~al., {Symmetry energy of dilute warm nuclear matter}, Phys.
  Rev. Lett. 104 (2010) 202501.
\newblock \href {http://arxiv.org/abs/1001.1102} {\path{arXiv:1001.1102}},
  \href {https://doi.org/10.1103/PhysRevLett.104.202501}
  {\path{doi:10.1103/PhysRevLett.104.202501}}.

\bibitem{Hagel:2014wja}
K.~Hagel, J.~B. Natowitz, G.~R\"opke, {The equation of state and symmetry
  energy of low density nuclear matter}, Eur. Phys. J. A 50 (2014) 39.
\newblock \href {http://arxiv.org/abs/1401.2074} {\path{arXiv:1401.2074}},
  \href {https://doi.org/10.1140/epja/i2014-14039-4}
  {\path{doi:10.1140/epja/i2014-14039-4}}.

\bibitem{Kekelidze:2016hhw}
V.~D. Kekelidze, R.~Lednicky, V.~A. Matveev, I.~N. Meshkov, A.~S. Sorin, G.~V.
  Trubnikov, {Three stages of the NICA accelerator complex}, Eur. Phys. J. A
  52~(8) (2016) 211.
\newblock \href {https://doi.org/10.1140/epja/i2016-16211-2}
  {\path{doi:10.1140/epja/i2016-16211-2}}.

\bibitem{Ropke:2017dur}
G.~R\"opke, D.~Blaschke, Y.~B. Ivanov, I.~Karpenko, O.~V. Rogachevsky, H.~H.
  Wolter, {Medium effects on freeze-out of light clusters at NICA energies},
  Phys. Part. Nucl. Lett. 15~(3) (2018) 225--229.
\newblock \href {http://arxiv.org/abs/1712.07645} {\path{arXiv:1712.07645}},
  \href {https://doi.org/10.1134/S1547477118030159}
  {\path{doi:10.1134/S1547477118030159}}.

\bibitem{HotQCD:2019xnw}
H.~T. Ding, et~al., {Chiral Phase Transition Temperature in ( 2+1 )-Flavor
  QCD}, Phys. Rev. Lett. 123~(6) (2019) 062002.
\newblock \href {http://arxiv.org/abs/1903.04801} {\path{arXiv:1903.04801}},
  \href {https://doi.org/10.1103/PhysRevLett.123.062002}
  {\path{doi:10.1103/PhysRevLett.123.062002}}.

\bibitem{ALICE:2013mez}
B.~Abelev, et~al., {Centrality dependence of $\pi$, K, p production in Pb-Pb
  collisions at $\sqrt{s_{NN}}$ = 2.76 TeV}, Phys. Rev. C 88 (2013) 044910.
\newblock \href {http://arxiv.org/abs/1303.0737} {\path{arXiv:1303.0737}},
  \href {https://doi.org/10.1103/PhysRevC.88.044910}
  {\path{doi:10.1103/PhysRevC.88.044910}}.

\bibitem{ALICE:2013cdo}
B.~B. Abelev, et~al., {$K^0_S$ and $\Lambda$ production in Pb-Pb collisions at
  $\sqrt{s_{NN}}$ = 2.76 TeV}, Phys. Rev. Lett. 111 (2013) 222301.
\newblock \href {http://arxiv.org/abs/1307.5530} {\path{arXiv:1307.5530}},
  \href {https://doi.org/10.1103/PhysRevLett.111.222301}
  {\path{doi:10.1103/PhysRevLett.111.222301}}.

\bibitem{ALICE:2013xmt}
B.~B. Abelev, et~al., {Multi-strange baryon production at mid-rapidity in Pb-Pb
  collisions at $\sqrt{s_{NN}}$ = 2.76 TeV}, Phys. Lett. B 728 (2014) 216--227,
  [Erratum: Phys.Lett.B 734, 409--410 (2014)].
\newblock \href {http://arxiv.org/abs/1307.5543} {\path{arXiv:1307.5543}},
  \href {https://doi.org/10.1016/j.physletb.2014.05.052}
  {\path{doi:10.1016/j.physletb.2014.05.052}}.

\bibitem{ALICE:2014jbq}
B.~B. Abelev, et~al., {$K^*(892)^0$ and $\phi(1020)$ production in Pb-Pb
  collisions at $\sqrt{s{NN}}$ = 2.76 TeV}, Phys. Rev. C 91 (2015) 024609.
\newblock \href {http://arxiv.org/abs/1404.0495} {\path{arXiv:1404.0495}},
  \href {https://doi.org/10.1103/PhysRevC.91.024609}
  {\path{doi:10.1103/PhysRevC.91.024609}}.

\bibitem{NA49:2002pzu}
S.~V. Afanasiev, et~al., {Energy dependence of pion and kaon production in
  central Pb + Pb collisions}, Phys. Rev. C 66 (2002) 054902.
\newblock \href {http://arxiv.org/abs/nucl-ex/0205002}
  {\path{arXiv:nucl-ex/0205002}}, \href
  {https://doi.org/10.1103/PhysRevC.66.054902}
  {\path{doi:10.1103/PhysRevC.66.054902}}.

\bibitem{NA49:2006gaj}
C.~Alt, et~al., {Energy and centrality dependence of anti-p and p production
  and the anti-Lambda/anti-p ratio in Pb+Pb collisions between 20/A-GeV and
  158/A-Gev}, Phys. Rev. C 73 (2006) 044910.
\newblock \href {https://doi.org/10.1103/PhysRevC.73.044910}
  {\path{doi:10.1103/PhysRevC.73.044910}}.

\bibitem{NA49:2008ysv}
C.~Alt, et~al., {Energy dependence of Lambda and Xi production in central Pb+Pb
  collisions at A-20, A-30, A-40, A-80, and A-158 GeV measured at the CERN
  Super Proton Synchrotron}, Phys. Rev. C 78 (2008) 034918.
\newblock \href {http://arxiv.org/abs/0804.3770} {\path{arXiv:0804.3770}},
  \href {https://doi.org/10.1103/PhysRevC.78.034918}
  {\path{doi:10.1103/PhysRevC.78.034918}}.

\bibitem{NA49:2008goy}
C.~Alt, et~al., {Energy dependence of phi meson production in central Pb+Pb
  collisions at s(NN)**(1/2) = 6 to 17 GeV}, Phys. Rev. C 78 (2008) 044907.
\newblock \href {http://arxiv.org/abs/0806.1937} {\path{arXiv:0806.1937}},
  \href {https://doi.org/10.1103/PhysRevC.78.044907}
  {\path{doi:10.1103/PhysRevC.78.044907}}.

\bibitem{NA49:2004irs}
C.~Alt, et~al., {Omega- and anti-Omega+ production in central Pb + Pb
  collisions at 40-AGeV and 158-AGeV}, Phys. Rev. Lett. 94 (2005) 192301.
\newblock \href {http://arxiv.org/abs/nucl-ex/0409004}
  {\path{arXiv:nucl-ex/0409004}}, \href
  {https://doi.org/10.1103/PhysRevLett.94.192301}
  {\path{doi:10.1103/PhysRevLett.94.192301}}.

\bibitem{NA49:2011blr}
T.~Anticic, et~al., {Antideuteron and deuteron production in mid-central Pb+Pb
  collisions at 158$A$ GeV}, Phys. Rev. C 85 (2012) 044913.
\newblock \href {http://arxiv.org/abs/1111.2588} {\path{arXiv:1111.2588}},
  \href {https://doi.org/10.1103/PhysRevC.85.044913}
  {\path{doi:10.1103/PhysRevC.85.044913}}.

\bibitem{Friese:2002re}
V.~Friese, {Production of strange resonances in C + C and Pb + Pb collisions at
  158-A-GeV}, Nucl. Phys. A 698 (2002) 487--490.
\newblock \href {https://doi.org/10.1016/S0375-9474(01)01410-5}
  {\path{doi:10.1016/S0375-9474(01)01410-5}}.

\bibitem{NA49:2011bfu}
T.~Anticic, et~al., {$K^{\ast}(892)^0$ and $\bar{K}^{\ast}(892)^0$ production
  in central Pb+Pb, Si+Si, C+C and inelastic p+p collisions at
  158$A$\textasciitilde{}GeV}, Phys. Rev. C 84 (2011) 064909.
\newblock \href {http://arxiv.org/abs/1105.3109} {\path{arXiv:1105.3109}},
  \href {https://doi.org/10.1103/PhysRevC.84.064909}
  {\path{doi:10.1103/PhysRevC.84.064909}}.

\bibitem{NA49:2007stj}
C.~Alt, et~al., {Pion and kaon production in central Pb + Pb collisions at 20-A
  and 30-A-GeV: Evidence for the onset of deconfinement}, Phys. Rev. C 77
  (2008) 024903.
\newblock \href {http://arxiv.org/abs/0710.0118} {\path{arXiv:0710.0118}},
  \href {https://doi.org/10.1103/PhysRevC.77.024903}
  {\path{doi:10.1103/PhysRevC.77.024903}}.

\bibitem{E-802:1999acy}
L.~Ahle, et~al., {Simultaneous multiplicity and forward energy characterization
  of particle spectra in Au + Au collisions at 11.6-A-GeV/c}, Phys. Rev. C 59
  (1999) 2173--2188.
\newblock \href {https://doi.org/10.1103/PhysRevC.59.2173}
  {\path{doi:10.1103/PhysRevC.59.2173}}.

\bibitem{E802:1999hit}
L.~Ahle, et~al., {Proton and deuteron production in Au + Au reactions at
  11.6/A-GeV/c}, Phys. Rev. C 60 (1999) 064901.
\newblock \href {https://doi.org/10.1103/PhysRevC.60.064901}
  {\path{doi:10.1103/PhysRevC.60.064901}}.

\bibitem{E-802:1999ewk}
L.~Ahle, et~al., {Centrality dependence of kaon yields in Si + A and Au + Au
  collisions at the AGS}, Phys. Rev. C 60 (1999) 044904.
\newblock \href {http://arxiv.org/abs/nucl-ex/9903009}
  {\path{arXiv:nucl-ex/9903009}}, \href
  {https://doi.org/10.1103/PhysRevC.60.044904}
  {\path{doi:10.1103/PhysRevC.60.044904}}.

\bibitem{Becattini:2000jw}
F.~Becattini, J.~Cleymans, A.~Keranen, E.~Suhonen, K.~Redlich, {Features of
  particle multiplicities and strangeness production in central heavy ion
  collisions between 1.7A-GeV/c and 158A-GeV/c}, Phys. Rev. C 64 (2001) 024901.
\newblock \href {http://arxiv.org/abs/hep-ph/0002267}
  {\path{arXiv:hep-ph/0002267}}, \href
  {https://doi.org/10.1103/PhysRevC.64.024901}
  {\path{doi:10.1103/PhysRevC.64.024901}}.

\bibitem{HADES:2015oef}
G.~Agakishiev, et~al., {Statistical hadronization model analysis of hadron
  yields in p + Nb and Ar + KCl at SIS18 energies}, Eur. Phys. J. A 52~(6)
  (2016) 178.
\newblock \href {http://arxiv.org/abs/1512.07070} {\path{arXiv:1512.07070}},
  \href {https://doi.org/10.1140/epja/i2016-16178-x}
  {\path{doi:10.1140/epja/i2016-16178-x}}.

\bibitem{Cleymans:1998yb}
J.~Cleymans, H.~Oeschler, K.~Redlich, {Influence of impact parameter on thermal
  description of relativistic heavy ion collisions at (1-2) A-GeV}, Phys. Rev.
  C 59 (1999) 1663.
\newblock \href {http://arxiv.org/abs/nucl-th/9809027}
  {\path{arXiv:nucl-th/9809027}}, \href
  {https://doi.org/10.1103/PhysRevC.59.1663}
  {\path{doi:10.1103/PhysRevC.59.1663}}.

\bibitem{Averbeck:2000sn}
R.~Averbeck, R.~Holzmann, V.~Metag, R.~S. Simon, {Neutral pions and eta mesons
  as probes of the hadronic fireball in nucleus-nucleus collisions around
  1-A-GeV}, Phys. Rev. C 67 (2003) 024903.
\newblock \href {http://arxiv.org/abs/nucl-ex/0012007}
  {\path{arXiv:nucl-ex/0012007}}, \href
  {https://doi.org/10.1103/PhysRevC.67.024903}
  {\path{doi:10.1103/PhysRevC.67.024903}}.

\bibitem{Hempel:2009mc}
M.~Hempel, J.~Schaffner-Bielich, {Statistical Model for a Complete Supernova
  Equation of State}, Nucl. Phys. A 837 (2010) 210--254.
\newblock \href {http://arxiv.org/abs/0911.4073} {\path{arXiv:0911.4073}},
  \href {https://doi.org/10.1016/j.nuclphysa.2010.02.010}
  {\path{doi:10.1016/j.nuclphysa.2010.02.010}}.

\bibitem{ParticleDataGroup:2020ssz}
P.~A. Zyla, et~al., {Review of Particle Physics}, PTEP 2020~(8) (2020) 083C01.
\newblock \href {https://doi.org/10.1093/ptep/ptaa104}
  {\path{doi:10.1093/ptep/ptaa104}}.

\bibitem{Albright:2014gva}
M.~Albright, J.~Kapusta, C.~Young, {Matching Excluded Volume Hadron Resonance
  Gas Models and Perturbative QCD to Lattice Calculations}, Phys. Rev. C 90~(2)
  (2014) 024915.
\newblock \href {http://arxiv.org/abs/1404.7540} {\path{arXiv:1404.7540}},
  \href {https://doi.org/10.1103/PhysRevC.90.024915}
  {\path{doi:10.1103/PhysRevC.90.024915}}.

\bibitem{Begun:2015ifa}
V.~Begun, W.~Florkowski, {Bose-Einstein condensation of pions in heavy-ion
  collisions at the CERN Large Hadron Collider (LHC) energies}, Phys. Rev. C 91
  (2015) 054909.
\newblock \href {http://arxiv.org/abs/1503.04040} {\path{arXiv:1503.04040}},
  \href {https://doi.org/10.1103/PhysRevC.91.054909}
  {\path{doi:10.1103/PhysRevC.91.054909}}.

\bibitem{Mir:2023wkm}
S.~A. Mir, N.~A. Rather, I.~M.~U. Din, S.~Uddin, {Hadron Production in
  Ultra-relativistic Nuclear Collisions and Finite Baryon-Size Effects}, arxiv
  (12 2023).
\newblock \href {http://arxiv.org/abs/2312.13079} {\path{arXiv:2312.13079}}.

\bibitem{Kahangirwe:2024xyl}
M.~Kahangirwe, I.~Gonzalez, J.~A. Mu\~noz, C.~Ratti, V.~Vovchenko, {Convergence
  properties of $T'$-Expansion Scheme: Hadron Resonance Gas and Cluster
  Expansion Model}, arxiv (8 2024).
\newblock \href {http://arxiv.org/abs/2408.04588} {\path{arXiv:2408.04588}}.

\bibitem{Randrup:2006nr}
J.~Randrup, J.~Cleymans, {Maximum freeze-out baryon density in nuclear
  collisions}, Phys. Rev. C 74 (2006) 047901.
\newblock \href {http://arxiv.org/abs/hep-ph/0607065}
  {\path{arXiv:hep-ph/0607065}}, \href
  {https://doi.org/10.1103/PhysRevC.74.047901}
  {\path{doi:10.1103/PhysRevC.74.047901}}.

\bibitem{Schulz:1983iho}
H.~Schulz, G.~R\"opke, K.~K. Gudima, V.~D. Toneev, {The coalescence phenomenon
  and the Pauli quenching in high-energy heavy ion collisions}, Phys. Lett. B
  124 (1983) 458--460.
\newblock \href {https://doi.org/10.1016/0370-2693(83)91550-2}
  {\path{doi:10.1016/0370-2693(83)91550-2}}.

\bibitem{Ropke:2011tr}
G.~R{\"o}pke, {Parametrization of light nuclei quasiparticle energy shifts and
  composition of warm and dense nuclear matter}, Nucl. Phys. A 867 (2011)
  66--80.
\newblock \href {http://arxiv.org/abs/1101.4685} {\path{arXiv:1101.4685}},
  \href {https://doi.org/10.1016/j.nuclphysa.2011.07.010}
  {\path{doi:10.1016/j.nuclphysa.2011.07.010}}.

\end{thebibliography}





\end{document}